\documentclass[12pt]{article}
\begin{document}

\title{Premetric electrodynamics\footnote{Accepted to {\it Advances in Applied Clifford Algebras}}}
\author{Bernard Jancewicz\\
Institute of Theoretical Physics, University of Wroc{\l}aw, \\
pl. Maksa Borna 9, PL-50-204 Wroc{\l}aw, Poland\\
{\sf email: bjan@ift.uni.wroc.pl}} \date{November 5, 2005}
\maketitle

\begin{abstract}
Classical electrodynamics can be divided into two parts. In the
first one, with the use of a plenty of directed quantities, namely
multivectors and differential forms, no scalar product is
necessary. It is called premetric electrodynamics. In this part,
principal laws of the theory can be tackled, among them the two
observer-independent Maxwell's equations. The second part concerns
specific media and requires establishing of a scalar product and,
consequently a metric. We present an axiomatic approach to
electrodynamics in which the metric enters as late as possible.
Also a line of research is mentioned in which the notion of
non-inertial observer is studied and its influence on
observer-dependent Maxwell equations.\\
MSC numbers 15A75, 53B50, 78A40
\end{abstract}

\section{Introduction}

In the last decades, a way of presenting electrodynamics has been
proposed based on a broad use of differential forms; see Refs.
\cite{Grauert}--\cite{Jancew1}. Within this approach, a question
was discussed whether electrodynamics can be formulated in such a
way that the metric of space-time doesn't enter the fundamental
laws of it which is called {\it premetric electrodynamics}, see
the papers \cite{Jancew1}--\cite{Oziew}. Related developments can
be also found in the books \cite{Lindell}--\cite{Delphe}. A
crowning achievement of this approach is the book by Hehl and
Obukhov \cite{Hehl}.

I consider useful to present this study to the Clifford Algebra
community. It is built deductively, i.e. in the form of axioms:
conservation of electric charge, magnetic flux and
energy-momentum. At the end, the metric is introduced by the
constitutive relations between the electromagnetic field
quantities, namely $D$, $H$, called {\it excitations}, united into
the four-dimensional quantity $G$, and $B$, $E$, called {\it field
strengths}, united into four-dimensional quantity $F$.

Each axiom introduces a physical quantity as a four-dimensional
differential form. In order to connect it with the known
three-dimensional objects, a local split of space-time $M$ into
space and time is needed. For this purpose, two notions are
needed, namely a scalar function $\sigma $ on $M$ and a family $L$
of curves. The level surfaces of $\sigma$, that is,
three-dimensional hypersurfaces $S$ (defined by condition $\sigma
=t=$ const) are spatial parts of the split. The family $L$ has the
following properties: (i) through each point of $M$ one curve
$\kappa $ from $L$ passes, (ii) two different curves do not
intersect, (iii) the curves are not tangent to hypersurfaces
$\sigma =$ const. These curves are time parts of the local
decomposition of the space-time in the following sense
\begin{equation} T_x(M)=T_x(\kappa )\oplus T_x(S), \label{split}
\end{equation}
 where $T_x$ denotes the tangent space at point $x\in M$. This
decomposition is not equivalent to introducing a metric on $M$,
since no scalar product is introduced on hypersurfaces $S$. The
decomposition (\ref{split}) allows only to claim that $T_x(S)$
{\it is transversal to} $T_x(\kappa )$.\footnote{It would be nice
if a Minkowski metric on M would be such that $T_x(S)$ is
orthogonal to $T_x(\kappa)$. This can be valid only for metrics
compatible with the given observer field, but it is not the most
general case, see Section 6.} The level surfaces of $\sigma $
determine also a measure of length on curves $\kappa $, i.e. a
time parameter $t$.

Let a curve $\kappa $ be given by the parametrization
$x_{\kappa}(t)$, then one can define vectors
$v=\hbox{d}x_{\kappa}/\hbox{d}t$. Since through each point of $M$
a curve from $L$ passes, we obtain in this manner a vector field
$v$ on $M$. It has the property that the value of the one-form
$d\sigma$ on vector $v$ is \footnote{In this formula and later,
$d$ denotes the four-dimensional exterior derivative.}
\begin{equation} d\sigma (v)=1. \label{sigma}\end{equation}
The pair $\sigma ,\,v$ satisfying (\ref{sigma}), after Cruz and
Oziewicz \cite{Oziew} can be named {\it observer field}.

The existence of the one-form $d\sigma $ and the vector field $v$
allows to introduce a $(k-1)$-form $\omega _1$ and a $k$-form
$\omega _2$ for an arbitrary $k$-form $\omega $ through the
formulas
\begin{equation} \omega _1=\omega
\lfloor v,~~~~\omega _2=\omega -\omega _1\wedge d\sigma .
\label{omega1}\end{equation}  The forms $\omega _i,~i=1,2$ are
parallel to the vector field $v$, because $\omega _i\lfloor v=0$.
The second equation (\ref{omega1}) can be rearranged so as to
obtain the following expression of the arbitrary $k$-form $\omega
$ by $\omega _i$'s:
\begin{equation} \omega =\omega _1\wedge d\sigma +\omega _2.
\label{omega}\end{equation}The right-hand side of (\ref{omega}) is
called by Cruz and Oziewicz \cite{Oziew} the {\it
observer-dependent decomposition} of $\omega$. Hehl and Obukhov
\cite{Hehl} call it {\it 1+3 decomposition}. The restrictions of
$\omega _i$ to the spatial surface $S$ can be called {\it
three-dimensional components} of the four-dimensional form $\omega
$.

\section{Axiom 1: electric charge conservation}

The conservation of electric charge was treated as a fundamental
law already in the time of Benjamin Franklin. Now one can catch
single electrons and protons in traps and count them individually,
hence we are sure that the matter carries as a {\it primary
quality} something called {\it electric charge} which can be
counted in principle.

The first axiom is most conveniently formulated in terms of
differential forms with the help of the twisted 3-form $J$ of {\it
electrical current density}. Its integral over a 3-dimensional
region yields the total charge contained in it which can be
determined by counting the charged particles. The conservation of
the charge can be described as the differential law:
\begin{equation} dJ=0. \label{charge}\end{equation}
This law is metric-independent since it is based on the {\it
counting} procedure of elementary charges.

By the prescription (\ref{omega1}) we introduce twisted 2-form $j$
and twisted 3-form $\rho $
\begin{equation} j=-J\lfloor v,~~~~\rho=J +j\wedge d\sigma .
\label{cur1}\end{equation} and 1+3 decompose the current density
$J$ according to (\ref{omega})
\begin{equation} J=-j\wedge d\sigma +\rho \label{cursplit}
\end{equation}
By restricting $j$ and $\rho $ to $S$, one obtains the
three-dimensional electric current density and the electric charge
density, respectively. The law (\ref{charge}) assumes the familiar
shape of the charge continuity equation
\begin{equation} \dot{\rho }+\underline{d}j=0, \label{chargecon}
\end{equation}
where the dot denotes\footnote{It is simultaneously the
Lie-\'{S}lebodzi\'{n}ski derivative with respect to $v$, namely
$\cal{L}$$_v\rho =d(v\rfloor \rho )+ v\rfloor(d\rho )$. See also
footnote in Section 4.} the partial derivative with respect to
parameter $t$ and $\underline{d}$ is the three-dimensional exterior
derivative acting on differential forms defined on $S$.

By a theorem of de Rham, equation (\ref{charge}) implies existence
of a twisted 2-form $G$ such that
\begin{equation} J=dG. \label{gauss}\end{equation}
In this manner the 2-form of electromagnetic excitation is
introduced. I propose to call it {\it Gauss field} in analogy to
the name Faraday field for a 2-form $F$, which will emerge later.
This field is not uniquely given by (\ref{gauss}), but it can be
determined in measurements with the aid of perfect conductors and
superconductors. The Gauss field can be 1+3 decomposed with the
use of (\ref{omega}):
\begin{equation} G=-H\wedge d\sigma +D \label{gaussplit}
\end{equation}
into the twisted 1-form $H=-G\lfloor v$ of the {\it magnetic
excitation} (traditionally: magnetic field) and the twisted 2-form
$D=G+H\wedge d\sigma $ of the {\it electric excitation} (traditional
name: electric displacement or electric induction). Then the
relation (\ref{gauss}) after restriction to $S$ splits into two
three-dimensional inhomogeneous Maxwell equations
\begin{equation} \underline{d}D=\rho ,~~~~~~~~~~-\dot{D}+
\underline{d}H=j. \label{inhmaxw} \end{equation}

\section{Axiom 2 and 3: Lorentz force density and magnetic
flux conservation}

Now we would like to introduce another electromagnetic quantity,
namely the Faraday field. In analogy to the electrostatic case the
simplest four-dimensional definition the electromagnetic field
strength reads:
$$\hbox{force~density} \sim \hbox{field~strength}\times
\hbox{charge~current~density}. $$ Since the force is a 1-form, the
force density should be a 1-form-valued 4-form. After introducing
a vector basis $e_{\beta }$ of the tangent space, we write down
the $\beta $-th component $f_{\beta }$ of the force density as the
4-form
\begin{equation} f_{\beta}=(e_{\beta }\rfloor F)\wedge J
\label{farad} \end{equation} This axiom can be treated as an
operational procedure for defining the {\it electromagnetic field
strength} 2-form $F$, the {\it Faraday field}. After introducing
the 1-form $E=F\lfloor v$ of the {\it electric field strength} and
the 2-form $B=F-E\wedge d\sigma $ of the {\it magnetic field
strength} (traditional name: magnetic induction), the 1+3
decomposition of $F$ reads
\begin{equation} F=E\wedge d\sigma +B, \label{faradsplit}
\end{equation}

All experiments done so far attest to the absence of magnetic
charge, hence we write the equation
\begin{equation}dF=0 \label{flux}\end{equation}
expressing the conservation of magnetic flux. The decomposition
(\ref{faradsplit}) applied to equation (\ref{flux}) and
restriction to $S$ splits it into two three-dimensional
homogeneous Maxwell equations:
\begin{equation} \underline{d}B=0,~~~~~~~~~~\dot{B}+
\underline{d}E=j. \label{hmaxw} \end{equation}

\section{Axiom 4: Energy-momentum density}

In electrodynamics, after formulating the Maxwell equations, one
has to specify the energy-momentum density of the electromagnetic
field. It should be a 1-form-valued 3-form. With the use of the
vector basis $e_{\beta }$ of the tangent space, we write down the
$\beta $-th component $\Sigma _{\beta }$ of the energy-momentum
density as the 3-form
\begin{equation} \Sigma _{\beta }=\frac{1}{2}[F\wedge (e_{\beta
}\rfloor G)-G\wedge (e_{\beta }\rfloor F)] \label{enmom}
\end{equation}
It can be checked that it has the following relation with the
Lorentz force density
\begin{equation} f_{\beta}=d\Sigma _{\beta }+X_{\beta} ,
\label{relat}\end{equation} where $X_{\beta }=\frac{1}{2}(G\wedge
\cal{L}_{\beta}F-F\wedge \cal{L}_{\beta}G)$ and $\cal{L}_{\beta}$
is the Lie-\'{S}lebodzi\'{n}ski derivative $\cal{L}_{\beta }\phi$
 $=d(e_{\beta }\rfloor \phi )+ e_{\beta }\rfloor(d\phi).
$\footnote{Sophus Lie (died in 1899) had not introduced any
derivative. This derivative was invented by W{\l}adys{\l}aw
\'{S}lebodzi\'{n}ski in 1931, see \cite{sle}. One year later, van
Dantzig \cite{dant} quoted the paper by \'{S}lebodzi\'{n}ski, but
named this notion Lie derivative.}

\section{Axiom 5: Constitutive relation}

Up to now all considerations are generally covariant and
metric-free. They are valid in flat Minkowskian as well as in
curved pseudo-Riemannian space-time. Therefore, the Maxwell's
equations represent the optimal formulation of classical
electrodynamics.

The electromagnetical theory in some moment incorporates the
metric of a flat or curved space-time via the constitutive
relation between the excitation and the field strength. In
particular, the standard Maxwell-Lorentz electrodynamics in the
vacuum arises when
\begin{equation} G=\lambda _0\star F. \label{constvac}\end{equation}
Here the Hodge star $\star $ ~is defined by the space-time metric
$g_{ij}=\hbox{diag}\,(c^2,-1,-1,-1)$, $\lambda
_0=\sqrt{\varepsilon _0/\mu_0}$ is the vacuum admittance.

For a general medium, a local and linear constitutive relation
seems to be a reasonable assumption of the axiomatic approach:
\begin{equation} G_{ij}=\frac{1}{2}\,\kappa _{ij}^{~~kl}F_{kl}.
\label{constrel}\end{equation} The constitutive tensor $\kappa
_{ij}^{~~kl}$, antisymmetric in pairs $ij$ and $kl$ separately,
has 36 independent components. It is useful to decompose it into
irreducible parts. Within the premetric framework, the contraction
is the only tool for this.

First contraction
\begin{equation} \kappa _i^{~k}=\kappa _{il}^{~~kl}
\label{contr1}\end{equation} has 16 components. Second contraction
\begin{equation} \kappa =\kappa _k^{~k}=\kappa _{kl}^{~~kl}\label{contr2}
\end{equation} is the single component. The traceless piece of $\kappa
_i^{~k}$:
\begin{equation}\pi _i^{~k}=\kappa _i^{~k}-\frac{1}{4}\,\kappa
\delta _i^k \label{contr3}\end{equation} has 15 components. The
$\kappa $ and $\pi $ can be immersed in the original constitutive
tensor as follows:
\begin{equation} \kappa _{ij}^{~~kl}=\omega _{ij}^{~~kl}+
2\pi _{[i}^{~[k} \delta _{j]}^{l]}+\frac{1}{6}\kappa \delta
_{[i}^{k}\delta _{j]}^{l} \label{const4}\end{equation} $\omega
_{.\,.}^{~~.\,.}$ is the totally traceless part of
$\kappa_{.\,.}^{~~.\,.}$:
\begin{equation} \omega _{il}^{~~kl}=0\label{const5} \end{equation}
The right-hand side of (\ref{const4}) is the split of $\kappa
_{.\,.}^{~~.\,.}$ according to numbers of components $36=20+15+1$.
$\omega _{.\,.}^{~~.\,.}$ is called {\it principal part} of the
constitutive tensor. $\pi _{.}^{~.}$, treated as a function of
space-time point, defines a {\it skewon field}:
\begin{equation} S_i^{~j}=-\frac{1}{2}\,\pi _i^{~j} \label{skewon}
\end{equation}
and $\kappa $ defines an {\it axion} field:
\begin{equation} \alpha =\frac{1}{12}\, \kappa . \label{axion}
\end{equation}
The standard Maxwell-Lorentz electrodynamics arises when $S=0$ and
$\alpha =0$, whereas
\begin{equation} \omega _{ij}^{~~~kl}=\lambda
_0\,\sqrt{-g}\,\epsilon _{ijmn}g^{mk}g^{nl}. \label{const6}
\end{equation}

It has been shown in \cite{lammer} that when electromagnetic waves
in the geometric optics approximation are to exist with the lack
of birefrigence in vacuum, this implies that there is only one
future and only one past directing light cone. As a consequence,
the signature of the metric is Lorentzian. The paper \cite{lammer}
has a nice continuation \cite{itin}.

It is worth to mention that even in flat space-time the spatial
part of the metric is not unique -- it can be accommodated to a
medium, especially when it is non-isotropic. For instance, when a
metric is introduced in which the electric permittivity tensor is
diagonal, all electrostatic problems can be reduced to that of the
isotropic dielectric, see \cite{Jancew1}. Similar observation is
valid for the anisotropic magnetic medium and the magnetostatics.

Recently an exotic medium was considered for which the principal
and skewon parts of the constitutive tensor are zero and only
axion part remains. In this case the constitutive relation assumes
the simplest form
\begin{equation} G=\alpha F \label{pemc} \end{equation}
with the pseudoscalar $\alpha $. Lindell and Sihvola \cite{lindel}
called it {\it perfect electromagnetic conductor (PEMC)}, because
the limit $\alpha \to 0$ yields the perfect magnetic conductor,
and the limit $1/\alpha \to 0$ gives the perfect electric
conductor. The proportionality relation (\ref{pemc}) implies that
the energy-momentum density (\ref{enmom}) is zero for the
electromagnetic field present in such a medium. This means that
when electromagnetic wave enters it, the wave can not transmit
energy. I have considered \cite{jancew3} plane electromagnetic
wave incident normally from an ordinary medium on a slab made of
PEMC. The boundary conditions on the front and back interface
between the PEMC and ordinary media on two sides imply that the
electromagnetic wave does not enter the second medium behind the
slab.

There is also another axiom in the book of Hehl and Obukhov,
namely that concerning the splitting of the electromagnetic
quantities in the continuous media into external and matter parts,
but I shall not dwell on it.

\section{Inertial versus non-inertial observers}

An interesting direction of research concerns the reference frames
and observers, see Refs. \cite{ekart}, \cite{demian}, \cite{fecko},
\cite{Kocik}, \cite{Oziew}, \cite{Hehl}, \cite{rod}. There is no
established terminology what is an observer and what is a reference
frame. For instance W.A. Rodrigues and E.C. de Oliveira \cite{rod}
define reference frame as as a (time-like) vector field and
observers as integral lines of the vector field. F.W. Hehl and Yu.N.
Obukhov \cite{Hehl} use exchangeably coordinate frame and frame of
reference. Most often an inertial reference frame is defined with
the aid of connection, see \cite{ekart}, \cite{demian} and
\cite{rod} Section 5.1.1. J.J. Cruz and Z. Oziewicz \cite{Oziew}
(see also \cite{fecko}, \cite{Kocik}) consider more general notion
of {\it observer field}, when the pair: one-form $d\sigma $ and
vector field $v$ of the Introduction are replaced by a (1,1)-tensor
field $p=v\otimes d\sigma $ which in each point of $M$ is a
projection of arbitrary tensors onto time axis of the observer. Then
they ponder on how one can discriminate between inertial and
non-inertial observers. Cruz and Oziewicz postulate that it is
possible to define an inertial observer without connection if one
introduces so called Fr\"{o}licher-Nijenhuis operation on
differential forms and Frobenius algebra of massive observers. Then
the observer is inertial when the vector field $v$ is
divergence-free and acceleration of the observer vanishes. (For the
exact meaning of these terms see \cite{Oziew}.)

The named authors consider also another generalization, namely
that the spaces $T_x(\kappa )$ and $T_x(S)$ are not orthogonal.
This means that the decomposition (\ref{split}) is not compatible
with the Minkowski-Riemann metric $g$ in $M$. In this case Cruz
and Oziewicz say that the observer field $p$ {\it is not
$g$-orthogonal}. They are also able to write the charge continuity
equation (\ref{chargecon}) and the four Maxwell equations
(\ref{inhmaxw}, \ref{hmaxw}) not by the restriction to spatial
surfaces $S$, but as equations for bona fide four-dimensional
observer-dependent differential forms $j,\,\rho ,\,E,\,B,\,H,$ and
$D$, and these equations are observer-dependent, but still
pre-metric.

Afterwards, the observer-dependent vector fields {\bf B}, {\bf E},
{\bf D}, {\bf H} are introduced in the presence of the space-time
metric $g$ and the Maxwell equations are written for them. In the
equations, extra terms are present when the observer field is
non-inertial. These terms depend on acceleration and rotary motion
of the observers. Some of these formulas were derived already by
Jerzy Kocik (\cite{Kocik}). The modified Maxwell equations are
displayed also in the book by Hehl and Obukhov, \cite{Hehl} Section
B.4.4, without introducing the vector fields. The additional terms
appear when the components of the electromagnetic field are
expressed in terms of the coframe of the non-inertial reference
frame. It is interesting to know about possible observability of
the extra terms. The Earth is an example of non-inertial system of
reference, hence such terms should be present in the Maxwell
equations in this system. A private message from Zbigniew
Oziewicz: the magnitude of them is too small to be detectable now.

\vspace{6mm} {\bf Acknowledgements}

\medskip
I am grateful to Friedrich Hehl and Zbigniew Oziewicz for numerous
explanations which helped to improve the article.


\begin{thebibliography}{99}

\bibitem{Grauert} H. Grauert and I. Lieb: {\it Differential und Integralrechnung},
vol~3, Springer Verlag, Berlin 1968.

\bibitem{Misner}  Charles~Misner, Kip~Thorne and John~Archibald~Wheeler: {\it
Gravitation}, Freeman and Co., San Francisco 1973, Sec. 2.5.

\bibitem{Thirring}  Walther Thirring: {\it Course in Mathematical Physics}, vol.
2: {\it Classical Field Theory}, Springer Verlag, New York 1979.

\bibitem{Deschamp}  G.A. Deschamps: ``Electromagnetics and differential forms'',
{\it Proc. IEEE} {\bf 69}(1981)676.

\bibitem{Schouten}  Jan~Arnoldus~Schouten: {\it Tensor Analysis for Physicists,}
Dover Publ., New York 1989 (first edition: Clarendon Press, Oxford 1951).

\bibitem{Frankel} Theodore Frankel: {\it Gravitational Curvature. An Introduction
to Einstein's Theory,} Freeman and Co., San Francisco 1979.

\bibitem{Burke1}  William L. Burke: {\it Applied Differential Geometry},
Cambridge University Press, Cambrigde 1985.

\bibitem{Ingarden}  Roman Ingarden and Andrzej Jamio{\l}kowski: {\it Classical
Electrodynamics}, El\-sie\-vier, Amsterdam 1985.

\bibitem{Meetz} Kurt Meetz, Walter L. Engel: {\it Elektromagnetische Felder},
Springer Verlag, Berlin 1980.

\bibitem{Burke2}  William L. Burke: {\it Spacetime, Geometry, Cosmology},
University Science Books, Mill Valley 1980.

\bibitem{Jancew1} Bernard Jancewicz: ``A variable metric electrodynamics. The
Coulomb and Biot-Savart laws in anisotropic media", {\it Ann.~Phys
(NY)} {\bf 245},2(1996)227.

\bibitem{True} C. Truesdell and R.A Toupin: ``The classical field theories",
in {\it Handbuch der Physik}, vol.III/1, S. Fl\"{u}gge, editor,
Springer Verlag, Berlin 1960, pp. 226-793.

\bibitem{Toupin} R.A. Toupin: ``Elasticity and electro-magnetics", in:
{\it Non-Linear Continuum Theories. CIME Conference, Bressanone,
Italy 1965}, C. Tuesdell and G. Grioli, coordinators, pp. 203-342.

\bibitem{Post1} E.J. Post {\it Formal Structure of Electromagnetics},
North-Holland, Amsterdam 1962, and Dover, New York 1997.

\bibitem{Post2} E.J. Post: ``The constitutive map and some of ramifications",
{\it Ann. Phys. (NY)} {\bf 71} (1972) 497-518.

\bibitem{Kiehn} R.M Kiehn, G.P. Kiehn and J.B. Roberts: ``Parity and
time-reversal symmetry breaking, singular solutions and Fresnel
surfaces" {\it Phys. Rev.} {\bf A43} (1991) 5665-5671.

\bibitem{Kovetz} A. Kovetz: {\it Electromagnetic Theory}, Oxford Univ. Press,
Oxford 2000.

\bibitem{Oziew} Jos\'{e} de Jesus Cruz Guzman and Zbigniew
Oziewicz: ``Fr\"{o}licher-Nijenhuis algebra and four Maxwell's
equations for non-inertial observer" {\it Bulletin de la
Soci\'{e}t\'{e} de Sciences et de Lettres de {\L}\'{o}d\'{z},
vol.} {\bf 53}, {\it S\'{e}rie Recherches sur les Deformations}
{\bf 39}(2003)107-160.

\bibitem{Lindell} I.V. Lindell: {\it Differential Forms in Electromagnetics},
IEEE Press, Piscataway NJ and Wiley-Interscience, 2004.

\bibitem{Russer} P. Russer: {\it Electromagnetics, Microwave Circuit and
Antenna Design for Communications Engineering}, Artech House, Boston 2003.

\bibitem{Delphe} D.H. Delphenich: ``On the axioms of topological
electromagnetism", {\it Ann. Phys. (Leipzig)} {\bf
14},6(2005)347-377.

\bibitem{Hehl} F.W. Hehl and Yu.N. Obukhov: {\it Foundations of
Classical Electrodynamics: Charge, Flux and Metric}.
Birkh\"{a}user, Boston 2003.

\bibitem{lammer} C. L\"{a}mmerzahl and F.W. Hehl: ``Riemannian
light cone from vanishing birefrigence in premetric vacuum
electrodynamics", {\it Phys. Rev.} {\bf D70} (2004) 1050022.

\bibitem{itin} Y. Itin: ``No-birefrigence conditions for
space-time", http://arXiv.org/hep-th/0508144.

\bibitem{sle} W{\l}adys{\l}aw \'{S}lebodzi\'{n}ski: ``Sur les
equations de Hamilton", {\it Bulletin de l'Acad\`{e}mie Royale
Belgique} {\bf 17},5(1931)865-870.

\bibitem{dant} D. van Dantzig: ``Zur allgemeinen projektiven
Differentialgeometrie", {\it Kon. Ak. v. Wetensch. Amsterdam}
 {\bf 35}(1932)524-542.

\bibitem{lindel} I.V. Lindell and A.H. Sihvola: ``Perfect
electromagnetic conductor", {\it J. Electrom. Wav. Appl.} {\bf
19},7(2005)861-869.

\bibitem{jancew3} Bernard Jancewicz: ``Plane electromagnetic wave
in PEMC", {\it J. Electrom. Wav. Appl} {\bf 20},5(2006)647-659,
arXiv:physics/0508231.

\bibitem{ekart} Carl Eckart: ``The thermodynamics of irreverisble processes.
Part III: Relativistic theory of the simple fluid", {\it Phys.
Rev.} {\bf 58}(1940)919-924.

\bibitem{demian} Marek Demia\'{n}ski: {\it Relativistic Astrophysics}, Pegamon
Press and PWN, Oxford and Warsaw 1985. (First Polish edition: Warsaw 1978.)

\bibitem{fecko} Marian Fecko: ``On 3+1 decompositions with respect
to an observer field via differential forms", {\it J. Math. Phys.}
{\bf 38},9(1997)4542-4560.

\bibitem{Kocik} Jerzy Kocik: ``Relativistic observer and
Maxwell's equations: an example of a nonprincipal Ehresmann
connection", http://www.math.siu.edu/Kocik,  1997.

\bibitem{rod} Waldyr Alves Rodrigues Jr and Edmundo Capelas de Oliveira:
{\it The many faces of Maxwell, Dirac and Einstein equations},\\
http://www.ime.unicamp.br/rel$\_$pesq/2005/relatorio05.html
\end{thebibliography}
\end{document}